\documentclass[graybox]{svmult}

\newcommand{\Mo}{\mbox{M$_{\odot}$}}
\newcommand{\HH}{\mbox{H$_2$}}
\newcommand{\kms}{\mbox{km~s$^{-1}$}}
\newcommand{\AD}{\mbox{ATLAS$^{\rm 3D}$}}

\def\rf@jnl#1{{#1}}
\def\aj{\rf@jnl{AJ }}                   
\def\araa{\rf@jnl{ARA\&A }}             
\def\aap{\rf@jnl{A\&A }}                
\def\apj{\rf@jnl{ApJ }}                 
\def\apjl{\rf@jnl{ApJ }}                
\def\apjs{\rf@jnl{ApJS }}               
\def\mnras{\rf@jnl{MNRAS }}             
\def\nat{\rf@jnl{Nature }}              
\def\iaucirc{\rf@jnl{IAU~Circ.}}       
\def\aplett{\rf@jnl{Astrophys.~Lett.}} 
\def\pasp{\rf@jnl{PASP }}               
\def\pasj{\rf@jnl{PASJ }}               

\usepackage{mathptmx}       
\usepackage{helvet}         
\usepackage{courier}        
\usepackage{type1cm}        
\usepackage{makeidx}         
\usepackage{graphicx}        
\usepackage{multicol}        
\usepackage[bottom]{footmisc}

\makeindex             

\begin{document}

\title*{Birth, life and survival of Tidal Dwarf Galaxies}
\author{Pierre-Alain Duc}
\institute{Pierre-Alain Duc \at Laboratoire AIM Paris Saclay \email{pierre-alain.duc@cea.fr}}
%
%
\maketitle

\abstract*{Advances on the formation and survival  of the so--called Tidal Dwarf Galaxies (TDGs) are reviewed.  The understanding  on how objects of the mass of dwarf galaxies  may form  in debris of galactic collisions has recently benefited from  the  coupling of multi-wavelength observations with  numerical simulations of galaxy mergers. Nonetheless, no consensual scenario has  yet emerged and as a matter of fact the very definition of TDGs remains elusive.  Their real cosmological importance is also a matter of debate,  their presence in our Local Group of galaxies  as well. Identifying old, evolved, TDGs among the population of regular dwarf galaxies and satellites may not be straightforward. However a number of specific properties (location, dark matter and  metal content) that objects of tidal origin should have are reminded here. Examples of newly discovered genuine old TDGs around a nearby elliptical galaxy    are finally presented. }

\abstract{Advances on the formation and survival  of the so--called Tidal Dwarf Galaxies (TDGs) are reviewed.  The understanding  on how objects of the mass of dwarf galaxies  may form  in debris of galactic collisions has recently benefited from  the  coupling of multi-wavelength observations with  numerical simulations of galaxy mergers. Nonetheless, no consensual scenario has  yet emerged and as a matter of fact the very definition of TDGs remains elusive.  Their real cosmological importance is also a matter of debate,  their presence in our Local Group of galaxies as well. Identifying old, evolved, TDGs among the population of regular dwarf galaxies and satellites may not be straightforward. However a number of specific properties (location, dark matter and  metal content) that objects of tidal origin should have are reminded here. Examples of newly discovered genuine old TDGs around a nearby elliptical galaxy    are finally presented. }

\section{Introducing Tidal Dwarf Galaxies}

There is yet no consensual definition of a {\em Tidal Dwarf Galaxy} (TDG). Let's however  stick to the acronym to define it:\\
$\bullet$  {\em Tidal} refers to an object made  of material that was tidally expelled from galaxies. The definition may be a bit enlarged and includes objects born in general   in debris of galaxy-galaxy collisions: tidal tails of mergers, but also collisional rings  and even gas stripped in the intergalactic medium by other processes than tidal forces. 
The key element is that the building material of TDGs used to belong to a larger parent  galaxy and was thus pre-enriched. In practice late--type, rotating colliding galaxies generate more debris than early--type, dynamically hot galaxies. As a consequence,  TDGs should mostly be  produced by wet mergers involving  spiral galaxies.  \\
$\bullet$    {\em Dwarf} means that the object born in tidal tails should have  the size and mass of a dwarf galaxy -- although this is a loose criterion given the large range of sizes/masses exhibited by dwarf galaxies (from the ultra faint ones around the Milky Way to the Magellanic type objects).  This specification allows to disentangle TDGs from the Super Star Clusters and other compact stellar objects that are formed as well in colliding systems. \\
$\bullet$    {\em Galaxy} implies that the system is  kinematically decoupled from its parent galaxy and gravitationally bound. This increases its ability to survive external gravitational  stirring or internal destructive processes such as stellar feedback. In other words, TDGs are not transient objects but correspond to genuine condensations of matter that have collapsed in-situ within collisional debris.\\ 
 
How does this definition of a TDG translate  into observational properties? \\
$\bullet$  Being recycled objects, TDGs have inherited from their parents the metal content of their  interstellar medium. Thus their metallicity  tells about the past chemical enrichment of their parents, and is thus not correlated with their actual mass, contrary to conventional galaxies. Made out of pre-enriched material, they should have an excess of heavy elements, provided that their parents were themselves metal rich. This implies as well that their dust content and molecular gas content, as traced by CO, is higher than in regular star--forming dwarf galaxies.\\
$\bullet$   Made out of material expelled from the dark--matter poor disks of their parent galaxies, TDGs have accreted little of their dark--matter content. As a consequence, their luminous mass (stars and gas) should be close to their dynamical, total, mass, contrary to conventional dark--matter dominated galaxies (but see Section~\ref{sec:2}). \\

\begin{figure}[t]
\includegraphics[width=\textwidth]{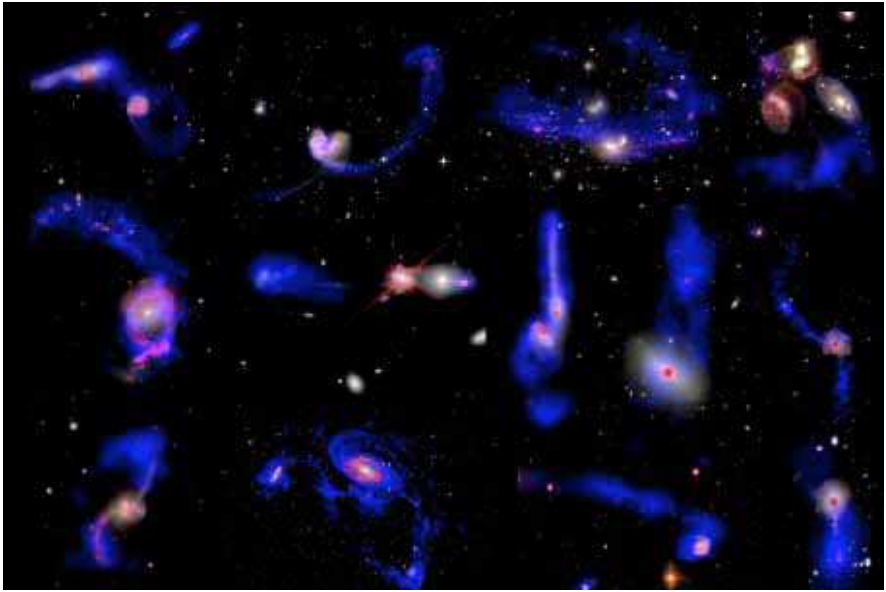}
%
%
\caption{Sample of colliding systems exhibiting TDG candidates. The distribution of the gas is shown in blue and the star-forming regions in red.}
\label{fig:tdg-obs}       
\end{figure}

 Examples of  observed Tidal Dwarf Galaxies are shown in Fig.~\ref{fig:tdg-obs}. On these images of colliding systems, the TDGs appear as red stains on blue ribbons, i.e. star--forming objects within gas--rich tails. The most massive of them are usually located near their tip.  
 Several papers have exploited the rich Ultraviolet/GALEX Infrared/Spitzer databases on interacting galaxies and investigated in details how star--formation proceeds in collisional debris  (e.g. \cite{Torres-Flores09,Smith10, Boquien10}).

 At this stage, it is worthwhile noting that the vast majority of the Tidal Dwarf Galaxies so far  securely  identified are young objects, formed in mergers that occurred less than  one Gyr ago. They still exhibit the umbilical cord linking them to their parents... i.e. the  tails and bridges in which they were formed have not had the time to evaporate. 
 Once evolved, TDGs should become  undistinguishable from regular satellite galaxies on optical images. 

\section{Birth of Tidal Dwarf Galaxies: models, simulations}
\label{sec:1}

Observations give some clues on the formation mechanism of tidal dwarfs. In the young TDGs observed so far, the atomic hydrogen makes the bulk of their mass. Therefore gas should play a key role.  On the theoretical side, several scenarios have been proposed, supported by various types of numerical simulations:\\
$\bullet$ (1)  {\it Local gravitational instabilities in the stellar component}. Simulations of mergers which only include the stellar component are apparently able to produce along tidal tails gravitational bound stellar objects, some reaching the  mass of dwarf galaxies   \cite{Barnes92}. However it has been claimed that they  might in fact be artifacts  of the N-body  simulations \cite{Wetzstein07} \\
$\bullet$ (2)  {\it Local gravitational instabilities in the gaseous component}.   In simulations which include the gas component,  real massive gas condensations may locally grow in the tails and form objects similar to TDGs  \cite{Wetzstein07} \\
$\bullet$  (3) {\it   Ejection of Jeans-unstable gas clouds}. Due to the increased velocity dispersion induced by galaxy-galaxy interactions, the Jeans mass of the individual cloud complexes increases in the outer disks of the parent galaxies. They are then pulled out by tidal forces, become unstable and  collapse when reaching large galacto-centric distances  \cite{Elmegreen93}.\\
$\bullet$  (4) {\it   A  top--down kinematical scenario}.    The {\it global} tidal field of galaxies with extended dark matter halos can efficiently carry  away from their disk a large fraction of the gas, while maintaining its surface density to a high value  \cite{Duc04b}. In fact tidal forces contribute to stretch the gas only at low galacto-centric distances, i.e. at the base of the tail.  As a result, gas accumulates near the tip of tidal tail, and then collapses and fragments, through a  process apparently opposite to  the  bottom--up one favored with the Cold Dark Matter model for the building up of classical galaxies.\\
$\bullet$  (5) {\it  The fully compressive mode of tidal forces}.    At locations where tidal forces are compressive rather than destructive, star / cluster formation may be triggered and/or already formed stellar objects, such as TDGs, may be protected from disruption \cite{Renaud09}.\\
$\bullet$   (6) {\it  Merger between Super-Star-Clusters}. SSCs with a range of masses may be formed in mergers.   Some of them might merge to reach the mass of dwarf galaxies \cite{Fellhauer02}. The TDGs born that way  would then resemble  the Ultra Compact Dwarf Galaxies (UCDs) identified in nearby groups and clusters of galaxies.\\

\begin{figure}[t]
\includegraphics[width=\textwidth]{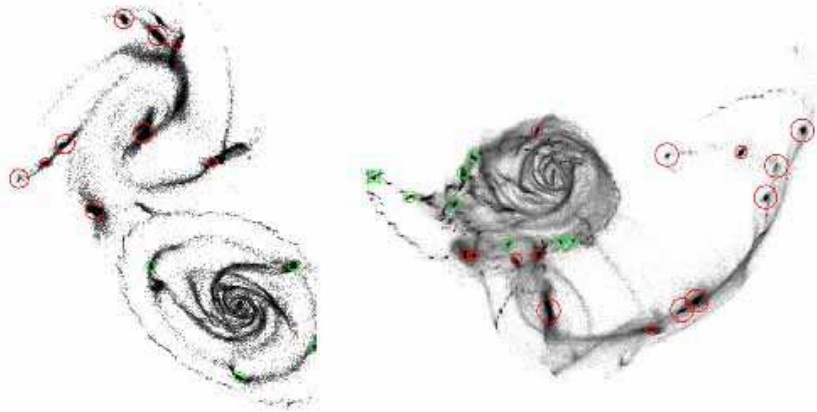}
%
%
\caption{Formation of tidal dwarf galaxies in high resolution numerical simulation of a major merger \cite{Bournaud08}. Two snapshots are shown, resp. after the first encounter and the merger (Belles et al., in prep) }
\label{fig:tdg-sim}       
\end{figure}

The variety of proposed scenarios tells how much  having ad-hoc  initial conditions and all necessary ingredients in the simulations is important:
 if the dark matter halo is truncated in the numerical simulations (to lower their computational cost), scenario (4) will not work; scenarios (2)-(5) require proper treatment of the gaseous component, including feedback. Scenario (6) needs high resolution, so as to resolve Super Star Clusters. 
  Fig.~\ref{fig:tdg-sim} presents one of such simulations fulfilling most of  these criteria. The numerical model used a total of 36 million particles, including 12 million ``sticky" particles for the gas component, and  minimal grid cell size of 32 pc  \cite{Bournaud08}. The production of  star clusters with masses down to $10^5~\Mo$ was directly resolved in these simulations. The mass spectrum of objects produced during the merger seems to be  bimodal. Two distinct families arise: (a)  compact  SSCs, with masses less than $10^8~\Mo$ which seem pressure supported and may be the progenitors of globular clusters; 
   (b) extended objects with masses above $10^8~\Mo$ which are usually supported by rotation. The latter have the properties of observed Tidal Dwarf Galaxies. Thus TDGs are not simply the high mass end of SSCs, a conclusion that was also reached from the analysis of HST images  \cite{Knierman03}. Furthermore, analyzing snapshots of the simulation for a period of one Gyr,  we found no evidence  that the latter evolve into the former, via merging.  The TDG progenitors are visible soon after the first encounter, in the outskirts of the colliding galaxies, at a time when the tidal tails have not yet completely unfolded. They quickly collect all  their building material. After about 100 Myr, their mass is stabilized. Rotational support appears as well very early on. Only a few massive objects are formed later on within the tidal tails.   
Note however that star-formation and gas feedback are not properly handled with the sticky particles used in these simulations. Investigations may now be carried using  fully hydrodynamical simulations   \cite{Teyssier10}.

\section{Life  of Tidal Dwarf Galaxies: dynamics}
\label{sec:2}
\begin{figure}
\includegraphics[width=\textwidth]{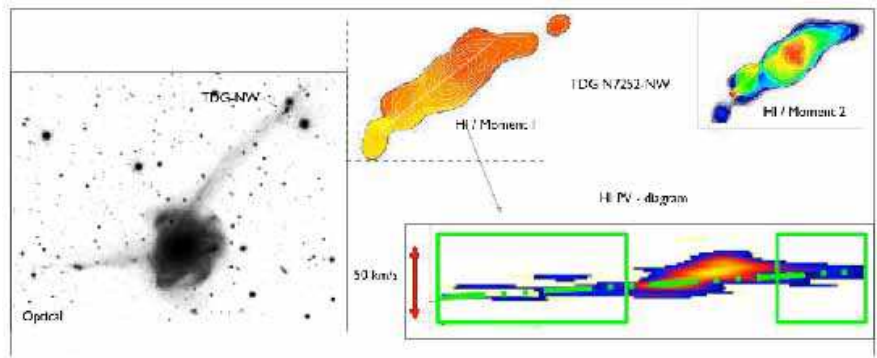}
%
%
\caption{Prototypical merger NGC 7252. The moment maps of the HI towards the TDG candidate (top), as well as its position--velocity diagrams are consistent with rotation (Belles et al. in prep.)  }
\label{fig:N7252}       
\end{figure}

Numerical models predict that TDGs, or at least the more massive of them, should be supported by rotation. Optical slit spectroscopy of numerous TDG candidates has been carried out   \cite{Weilbacher02,Mendes04}, showing strong velocity gradients, consistent with rotation, but also with artificial slit effects \cite{Weilbacher07}.
Integral field spectroscopy data \cite{Bournaud04}, as well as high resolution HI and CO datacubes \cite{Hibbard01,Bournaud07} are available for several TDGs and confirm that TDGs rotate, though at velocities much lower than measured with slit spectroscopy. 
For the TDGs located  in the collisional ring of NGC~5291,  the data quality was sufficient to  allow a determination of their total mass, and comparing it with the luminous mass (HI, \HH, stars), of their dark matter content. As predicted by numerical simulations and early estimates of the dynamical  mass based on the millimeter CO line width \cite{Braine00}, the inferred M/L is much lower than in regular, dark matter dominated dwarf galaxies. However  an unanticipated  mismatch by a factor of 2--3 between the dynamical and luminous mass has been noticed \cite{Bournaud07}. A similar discrepancy was found  for the TDG candidate VCC~2062, in the Virgo cluster  \cite{Duc07b}, and more recently  in one of the TDGs hosted by the prototypical advanced merger NGC~7252. The results of this latter study are illustrated in Fig.~\ref{fig:N7252}.

Various hypotheses have been put forward to account for the missing mass in TDGs. Cosmological dark matter accreted from disrupted satellite galaxies \cite{Read08} might be present in the disk of spirals and thus in TDGs. Theory of modified gravity, such as MOND, predicts rotation curves for TDGs similar to the observed ones \cite{Gentile07,Milgrom07}. An alternative idea is that  spirals disks contain dark baryons, for instance in the form of very cold molecular gas not accounted for by CO observations \cite{Pfenniger94}. The presence of dark gas clouds in the Milky Way  had been inferred  using gamma rays \cite{Grenier05}. In the far-infrared domain, Planck is also making a census of the molecular component not mapped by standard tracers. Whether it may explain entirely the missing mass observed in TDGs is still an open question.

\section{Death of Tidal Dwarf Galaxies: life expectancy, census}
\label{sec:3}
If  Tidal Dwarf Galaxies tell something about dark matter and thus about cosmology are they cosmological important objects? To answer this question, one should determine how many of them are produced per merger, and then how many manage to survive. Both simulations and observations may give clues on these issues. 

\begin{figure}
\includegraphics[width=\textwidth]{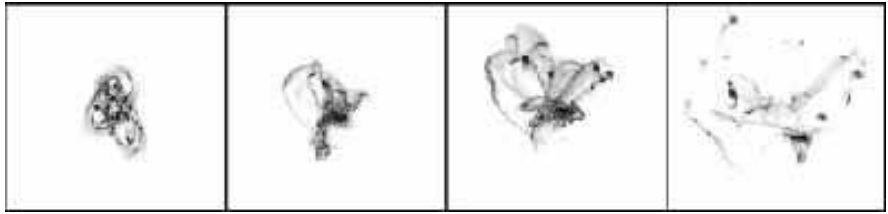}
%
%
\caption{TDG formation at high redshift, probed by simulations of clumpy disk galaxies with a high gas fraction (Bournaud et al., 2010, submitted)}
\label{fig:tdg-highz}       
\end{figure}

From the simple argument that, at high redshift, tidal collisions should have been numerous, some researchers reached the  radical conclusion that most dwarfs in the Universe should have a tidal origin \cite{Okazaki00}. Since this is rather unlikely, the Cold Dark Matter paradigm, and the hierarchical  mass assembly it implies have been questioned \cite{Kroupa10}.

 Analyzing a large set of numerical simulations,  we concluded that the formation of TDGs was in fact not a very efficient process in galaxy collisions: specific conditions should be met, such as low impact velocities, up to 250~\kms, leading to mergers, prograde encounters, mass ratios up to 4:1-- excluding minor mergers --, and above all initially extended gas in the parent galaxies  \cite{Bournaud06}. Furthermore, only TDGs located near the tip of the tidal tails are able to survive more than 1 Gyr.  The production rate is then  about 1 TDG per {\it favorable} merger.   Even if the merging rate increases with redshift, it would then be unlikely that TDGs contribute more than a few percent to the population of dwarf galaxies. However, the initial parameters of our simulations  \cite{Bournaud06}  are valid for nearby mergers. Simulations of mergers  tuned for the distant Universe which in particular assume that the disk of the parent galaxies had a higher gas fraction (up to 50\%) and was more turbulent, are presented in Fig. \ref{fig:tdg-highz}. They did not generate the very long tidal tails observed in nearby mergers. However a large number of clumps of matter, with typical masses of $10^8 - 10^9~\Mo$, initially formed in the disks may be kicked out by the collision.  Such objects, once independent, have all the properties expected for TDGs. 
 
 If such process is as efficient as these simulations show, the Local Universe should be full of such second generation, dark matter poor, dwarf galaxies. Is it really the case? The distribution of the Local Group dwarf spheroidals  on specific planes/circles   may suggest that they are old TDGs \cite{Metz07}. Given their location,  even the Magellanic Clouds were speculated to have been synthesized in an old merger that built  the present-day Andromeda galaxy \cite{Yang10}. However to validate such an hypothesis, one needs to check whether it is consistent with all the other properties expected for TDGs: lack of dark matter, specific star formation and chemical enrichment histories. And what is so far known about the properties of the neighbors of the Milky Way does not really support the tidal hypothesis.
 
\begin{figure}
\sidecaption
\includegraphics[width=7cm]{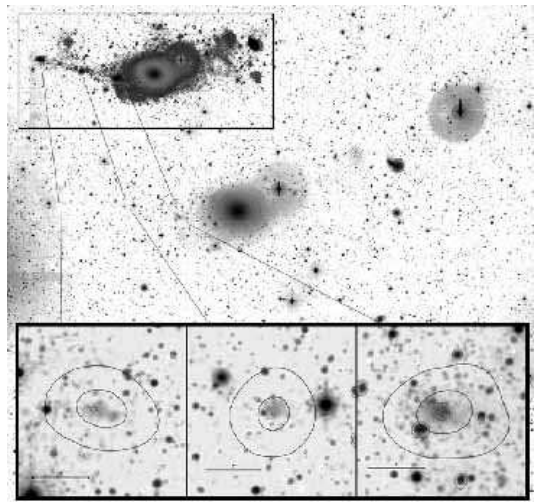}
%
%
\caption{Discovery of old TDGs around an  elliptical galaxy  with the CFHT MegaCam camera. The image was obtained as part of the \AD\ survey (Cappellari et al., 2011). The three candidates which have morphologies of dEs but are gas--rich -- see the bottom g--band images with contours of the HI emission from the WSRT (Serra et al., in prep.) superimposed --  lie along a 160~kpc long tidal tail, visible on the top image, where low surface brightness features have been enhanced. The merger has an estimated age of 2-5 Gyr (Duc et al., 2011, submitted)  }
\label{fig:N5557}       
\end{figure}

 Old TDGs remain to be found, first looking at environments where they are expected to have formed efficiently, i.e.  where major mergers have likely occurred. If Early-type galaxies result from major mergers, some of their satellites might be of tidal origin. Fig.~\ref{fig:N5557} presents deep optical images of a nearby elliptical galaxy that revealed the presence of three  gas--rich TDG candidates, which have likely formed 2-3 Gyr ago.
 
 Such systematic census of old TDGs, coupled with new numerical simulations of mergers, with a proper treatment of the gas and star-formation,  should be pursued to determine the real numerical importance of tidal dwarfs.

\begin{acknowledgement}
I wish to thank all my collaborators, observers and simulation experts, in particular Fr\'ed\'eric Bournaud, Pierre-Emmanuel Belles, M\'ed\'eric Boquien,  Elias Brinks, Ute Lisenfeld, Jonathan Braine, Peter Weilbacher, Etienne Ferriere and Daniel Miralles. The \AD\ team, in particular Paolo Serra,  is thanked for providing the data shown in Fig.~5.  Many thanks to Polis Papaderos for the initiative and  organization of this symposium.

\end{acknowledgement}
%



\begin{thebibliography}{10}
\providecommand{\url}[1]{{#1}}
\providecommand{\urlprefix}{URL }
\expandafter\ifx\csname urlstyle\endcsname\relax
  \providecommand{\doi}[1]{DOI~\discretionary{}{}{}#1}\else
  \providecommand{\doi}{DOI~\discretionary{}{}{}\begingroup
  \urlstyle{rm}\Url}\fi

\bibitem{Barnes92}
{Barnes}, J.E., {Hernquist}, L.: {Formation of dwarf galaxies in tidal tails}.
\newblock \nat \textbf{360}, 715--717 (1992)

\bibitem{Boquien10}
{Boquien}, M., {Duc}, P., {Galliano}, F., {Braine}, J., {Lisenfeld}, U.,
  {Charmandaris}, V., {Appleton}, P.N.: {Star Formation in Collision Debris:
  Insights from the modeling of their Spectral Energy Distribution}.
\newblock AJ \textbf{140}, 2124 (2010)

\bibitem{Bournaud06}
{Bournaud}, F., {Duc}, P.A.: {From tidal dwarf galaxies to satellite galaxies}.
\newblock \aap \textbf{456}, 481--492 (2006)

\bibitem{Bournaud04}
{Bournaud}, F., {Duc}, P.A., {Amram}, P., {Combes}, F., {Gach}, J.L.:
  {Kinematics of tidal tails in interacting galaxies: Tidal dwarf galaxies and
  projection effects}.
\newblock \aap \textbf{425}, 813--823 (2004)

\bibitem{Bournaud07}
{Bournaud}, F., {Duc}, P.A., {Brinks}, E., {Boquien}, M., {Amram}, P.,
  {Lisenfeld}, U., {Koribalski}, B.S., {Walter}, F., {Charmandaris}, V.:
  {Missing Mass in Collisional Debris from Galaxies}.
\newblock Science \textbf{316}, 1166  (2007)

\bibitem{Bournaud08}
{Bournaud}, F., {Duc}, P.A., {Emsellem}, E.: {High-resolution simulations of
  galaxy mergers: resolving globular cluster formation}.
\newblock \mnras \textbf{389}, L8--L12 (2008)

\bibitem{Braine00}
Braine, J., Lisenfeld, U., Duc, P.A., Leon, S.: Detection of molecular gas in
  tidal dwarf galaxies.
\newblock \nat \textbf{403}, 6772 (2000)

\bibitem{Cappellari11}
{Cappellari, M. et al.} 2011
\newblock \mnras in press (arXiv:1012.1551)


\bibitem{Duc04b}
{Duc}, P.A., {Bournaud}, F., {Masset}, F.: {A top-down scenario for the
  formation of massive Tidal Dwarf Galaxies}.
\newblock \aap \textbf{427}, 803--814 (2004)

\bibitem{Duc07b}
{Duc}, P.A., {Braine}, J., {Lisenfeld}, U., {Brinks}, E., {Boquien}, M.: {VCC
  2062: an old tidal dwarf galaxy in the Virgo cluster?}
\newblock \aap \textbf{475}, 187--197 (2007)

\bibitem{Elmegreen93}
Elmegreen, B.G., Kaufman, M., Thomasson, M.: An interaction model for the
  formation of dwarf galaxies and 10 exp 8 solar mass clouds in spiral disks.
\newblock ApJ \textbf{412}, 90 (1993)

\bibitem{Fellhauer02}
{Fellhauer}, M., {Kroupa}, P.: {The formation of ultracompact dwarf galaxies}.
\newblock \mnras \textbf{330}, 642--650 (2002)

\bibitem{Gentile07}
{Gentile}, G., {Famaey}, B., {Combes}, F., {Kroupa}, P., {Zhao}, H.S., {Tiret},
  O.: {Tidal dwarf galaxies as a test of fundamental physics}.
\newblock \aap \textbf{472}, L25--L28 (2007)

\bibitem{Grenier05}
{Grenier}, I.A., {Casandjian}, J.M., {Terrier}, R.: {Unveiling Extensive Clouds
  of Dark Gas in the Solar Neighborhood}.
\newblock Science \textbf{307}, 1292--1295 (2005)

\bibitem{Hibbard01}
{Hibbard}, J.E., {van der Hulst}, J.M., {Barnes}, J.E., {Rich}, R.M.:
  {High-Resolution H I Mapping of NGC 4038/39 (``The Antennae'') and Its Tidal
  Dwarf Galaxy Candidates}.
\newblock \aj \textbf{122}, 2969--2992 (2001)

\bibitem{Knierman03}
{Knierman}, K.A., {Gallagher}, S.C., {Charlton}, J.C., {Hunsberger}, S.D.,
  {Whitmore}, B., {Kundu}, A., {Hibbard}, J.E., {Zaritsky}, D.: {From Globular
  Clusters to Tidal Dwarfs: Structure Formation in the Tidal Tails of Merging
  Galaxies}.
\newblock \aj \textbf{126}, 1227--1244 (2003)

\bibitem{Kroupa10}
{Kroupa}, P., {Famaey}, B., {de Boer}, K.S., {Dabringhausen}, J., {Pawlowski},
  M.S., {Boily}, C.M., {Jerjen}, H., {Forbes}, D., {Hensler}, G., {Metz}, M.:
  {Local-Group tests of dark-matter concordance cosmology . Towards a new
  paradigm for structure formation}.
\newblock \aap \textbf{523}, A32 (2010)

\bibitem{Mendes04}
{Mendes de Oliveira}, C., {Cypriano}, E.S., {Sodr{\'e}}, L., {Balkowski}, C.:
  {A Nursery of Young Objects: Intergalactic H II Regions in Stephan's
  Quintet}.
\newblock \apjl \textbf{605}, L17--L20 (2004)

\bibitem{Metz07}
{Metz}, M., {Kroupa}, P.: {Dwarf spheroidal satellites: are they of tidal
  origin?}
\newblock \mnras \textbf{376}, 387--392 (2007)

\bibitem{Milgrom07}
{Milgrom}, M.: {MOND and the Mass Discrepancies in Tidal Dwarf Galaxies}.
\newblock \apjl \textbf{667}, L45--L48 (2007)

\bibitem{Okazaki00}
{Okazaki}, T., {Taniguchi}, Y.: {Dwarf Galaxy Formation Induced by Galaxy
  Interactions}.
\newblock \apj \textbf{543}, 149--152 (2000)

\bibitem{Pfenniger94}
{Pfenniger}, D., {Combes}, F., {Martinet}, L.: {Is dark matter in spiral
  galaxies cold gas? I. Observational constraints and dynamical clues about
  galaxy evolution}.
\newblock \aap \textbf{285}, 79--93 (1994)

\bibitem{Read08}
{Read}, J.I., {Lake}, G., {Agertz}, O., {Debattista}, V.P.: {Thin, thick and
  dark discs in {$\Lambda$}CDM}.
\newblock \mnras \textbf{389}, 1041--1057 (2008)

\bibitem{Renaud09}
{Renaud}, F., {Boily}, C.M., {Naab}, T., {Theis}, C.: {Fully Compressive Tides
  in Galaxy Mergers}.
\newblock \apj \textbf{706}, 67--82 (2009)

\bibitem{Smith10}
{Smith}, B.J., {Giroux}, M.L., {Struck}, C., {Hancock}, M., {Hurlock}, S.:
  {Spirals, Bridges, and Tails: A GALEX UV Atlas of Interacting Galaxies}.
\newblock AJ, \textbf{139}, 2719 (2010)

\bibitem{Teyssier10}
{Teyssier}, R., {Chapon}, D., {Bournaud}, F.: {The Driving Mechanism of
  Starbursts in Galaxy Mergers}.
\newblock \apjl \textbf{720}, L149--L154 (2010)

\bibitem{Torres-Flores09}
{Torres-Flores}, S., {Mendes de Oliveira}, C., {de Mello}, D.F., {Amram}, P.,
  {Plana}, H., {Epinat}, B., {Iglesias-P{\'a}ramo}, J.: {Star formation in the
  intragroup medium and other diagnostics of the evolutionary stages of compact
  groups of galaxies}.
\newblock \aap \textbf{507}, 723--746 (2009)

\bibitem{Weilbacher07}
{Weilbacher}, P.M., {Duc}, P.A.: {News from the ``Dentist's Chair'':
  Observations of AM1353-272 with the VIMOS IFU}.
\newblock In: M.~{Kissler-Patig}, J.R. {Walsh}, M.M. {Roth} (eds.) Science
  Perspectives for 3D Spectroscopy, pp. 207 (2007)

\bibitem{Weilbacher02}
{Weilbacher}, P.M., {Fritze-v.~Alvensleben}, U., {Duc}, P., {Fricke}, K.J.:
  {Large Velocity Gradients in the Tidal Tails of the Interacting Galaxy AM
  1353-272 (``The Dentist's Chair'')}.
\newblock \apjl \textbf{579}, L79--L82 (2002)

\bibitem{Wetzstein07}
{Wetzstein}, M., {Naab}, T., {Burkert}, A.: {Do dwarf galaxies form in tidal
  tails?}
\newblock \mnras \textbf{375}, 805--820 (2007)

\bibitem{Yang10}
{Yang}, Y., {Hammer}, F.: {Could the Magellanic Clouds be Tidal Dwarfs Expelled
  from a Past-merger Event Occurring in Andromeda?}
\newblock \apjl \textbf{725}, L24--L27 (2010)

\end{thebibliography}
\end{document}